\def\aj{AJ}
\def\araa{ARA\&A}
\def\apj{ApJ}
\def\apjl{ApJ}
\def\aap{A\&A}
 
\def\mnras{MNRAS}
\def\nat{Nature}

\newcommand{\emacss}{{\sc emacss}}
\newcommand{\feh}{{\rm [Fe/H]}}
\newcommand{\msun}{\,M_\odot}
\newcommand{\lsun}{\,L_\odot}
\newcommand{\ml}{M/L_V}
\newcommand{\mi}{M_{\rm i}}

\newcommand{\rh}{r_{\rm h}}
\newcommand{\rhi}{r_{\rm h,i}}
\newcommand{\rhdot}{\dot{r}_{\rm h}}
\newcommand{\rhoh}{\rho_{\rm h}}
\newcommand{\rhotid}{\rho_{\rm tid}}
\newcommand{\murho}{\mu_\rho}
\newcommand{\sigmarho}{\sigma_\rho}
\newcommand{\rg}{R_{\rm G}}
\newcommand{\rgi}{R_{\rm G,i}}
\newcommand{\pc}{{\rm pc}}

\newcommand{\thetapop}{\Theta^{\rm pop}}
\newcommand{\thetagc}{\Theta_j^{\rm GC}}

\newcommand{\gcdat}{X_j^{\rm GC}}
\newcommand{\popdat}{{\emph{\textbf X}}^{\rm pop}}
\newcommand{\ngc}{N_{\rm GC}}

\newcommand{\trh}{\tau_{\rm rh}}

\documentclass{iau} 
\usepackage{graphicx,natbib, amsmath, hyperref}

\title[IAUS 316.~~Inverting the dynamical evolution of GCs] 
{Inverting the dynamical evolution of \\ globular clusters: clues to their origin}

\author[Mark Gieles \& Poul Alexander]   
{Mark Gieles$^1$
 \and Poul Alexander$^{2}$}

\affiliation{$^1$Department of Physics, University of Surrey\\ 
 Guildford GU2 7XH, United Kingdom\\
email: {\tt m.gieles@surrey.ac.uk} \\[\affilskip]
$^2$Institute of Astronomy, University of Cambridge\\
Madingley Road, Cambridge CB3 0HA\\
email: {\tt poul.alexander@gmail.com}}

\pubyear{2015}
\volume{316}  
\jname{Formation, evolution, and survival of massive star clusters}
\editors{C. Charbonnel \& A. Nota, eds.}
\begin{document}

\maketitle

\begin{abstract}
Scaling relations for globular clusters (GCs) differ from scaling
relations for pressure supported (elliptical) galaxies. We show that
two-body relaxation is the dominant mechanism in shaping the bivariate
dependence of density on mass and Galactocentric distance for Milky Way GCs with
masses $\lesssim10^6\msun$, and it is possible, but not required, that GCs formed
with similar scaling relations as ultra-compact dwarf galaxies. We use
a fast cluster evolution model to fit a parameterised model for the
initial properties of Milky Way GCs to the observed present-day
properties. The best-fit cluster initial mass function is
substantially flatter (power-law index $\alpha=-0.6\pm0.2$) than what
is observed for young massive clusters (YMCs) forming in the nearby
Universe ($\alpha\simeq-2$). A slightly steeper CIMF is allowed when
considering the metal-rich GCs separately
($\alpha\simeq-1.2\pm0.4$). If stellar mass loss and two-body
relaxation in the Milky Way tidal field are the dominant disruption
mechanisms, then GCs formed differently from YMCs.

\keywords{globular clusters: general --
galaxies: star clusters --
Galaxy: formation --
stars: formation
}

\end{abstract}

\firstsection 
\section{Introduction}
Globular clusters (GCs) are a common component of almost every galaxy and their properties can be used to put constraints on the formation of their host galaxy. 
For example, GC ages and metallicities inform us about the assembly history of  galaxies \citep[e.g.][]{sz78, leaman13} and the spatial distribution of GCs within galaxies depends on the epoch of reinoization \citep{moore06, spitler12}. The similarity between structural properties  of young halo clusters in the Milky Way and clusters in external satellite galaxies such as the Magellanic Clouds, has been used to distinguish between accreted GCs and those that formed in-situ \citep[e.g.][]{mackey04, mackey05}. 

From a combination of data of nearby GCs and extra-galactic stellar systems \citep[e.g.][]{brodie11}, a break was found in the luminosity-size relation at a luminosity of a few $10^6\lsun$ \citep[e.g.][]{misgeld11}. 
Objects  above  this limit (massive star clusters and ultra-compact dwarf galaxies, UCDs) display a positive correlation between size and luminosity/mass \citep{kissler06}, following the extension of the relation of early type galaxies,  whereas the radii of fainter GCs are  independent of luminosity/mass, or slightly anti-correlate with luminosity/mass.  It is still an open question whether GCs and UCDs share a common formation mechanism, or evolutionary processes are responsible for the difference in structural properties  \citep{hilker07}. The half-mass relaxation timescale ($\trh$) of stellar systems with a mass of $10^6\msun$ and a radius of a few pc, is about a Hubble time and it is therefore plausible that  properties of GCs, i.e. systems older than a   $\trh$,  are affected by this process (\S\,\ref{sec:simple}). 

There are other indications that things change near $10^6\msun$: the mass-to-light ratio, $\ml$,  of GCs is about half that of UCDs \citep{mieske08}, which may point at the presence of dark matter in UCDs or variations in the stellar initial mass function. For at least one UCD it has been shown that a massive  black hole is responsible for  the elevated $\ml$ \citep{seth14}. But $\ml$ of stellar systems with short relaxation times can be underestimated because of a  mass segregation bias, if the assumption is made that light traces mass in deriving the dynamical mass \citep{shanahan15,sollima15}. Finally, if GCs formed embedded in a dark matter halo, it would have been pushed away by the stars as the result of dynamical friction \citep{baumgardt08}. Considering all of this, it is not clear that the difference in $\ml$ between GCs and UCDs points at a genuine difference in nature, or is the result of nurture.

\begin{figure}
\begin{center}
 \includegraphics[width=0.49\columnwidth]{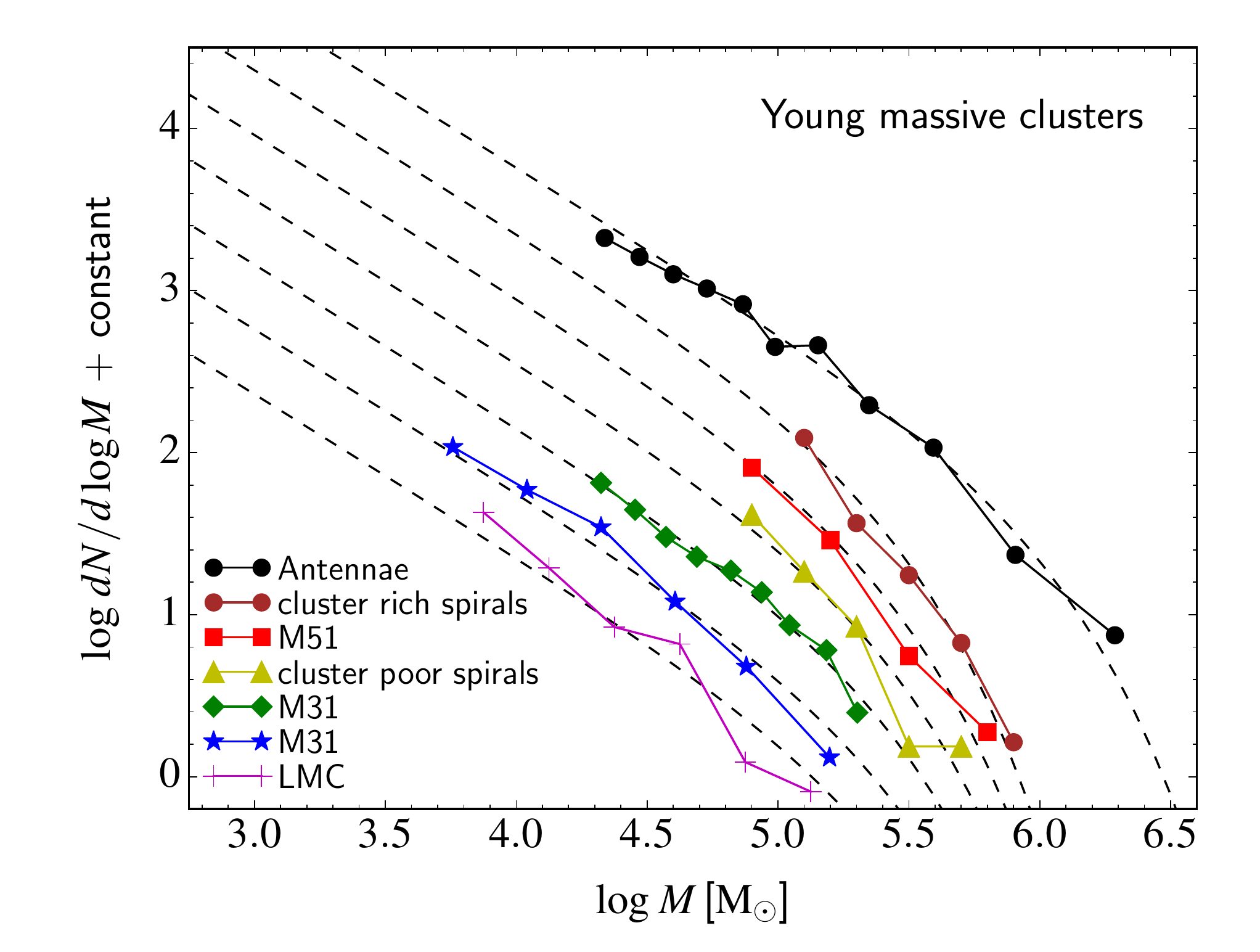} 
 \includegraphics[width=0.49\columnwidth]{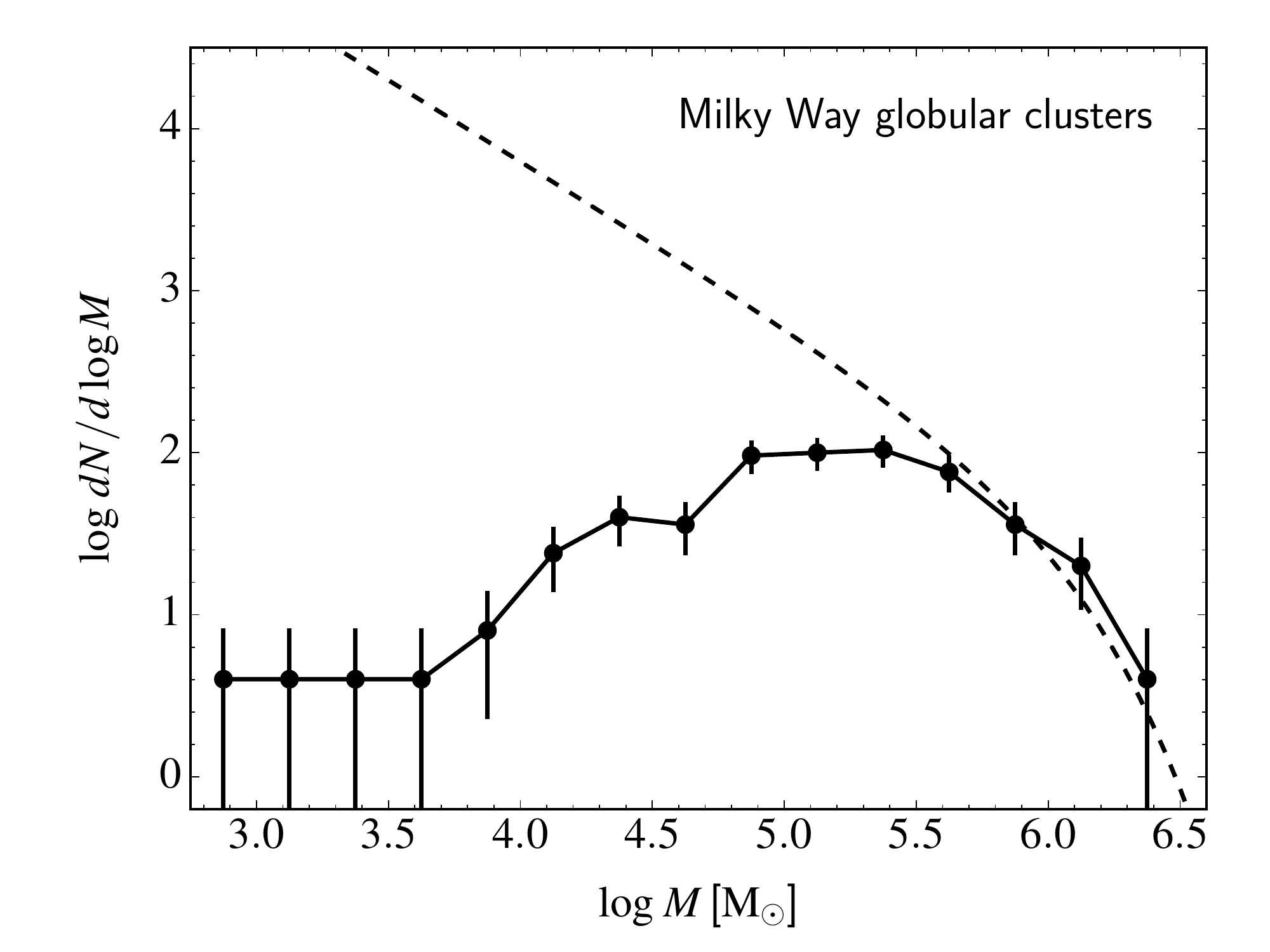} 
 \caption{{\it Left}: Mass distribution of YMCs are well described by $dN/dM \propto M^{-2}$ and an exponential cut-off at $M_* \simeq 10^5-10^6\msun$ (data from \citealt*{portegies10}). {\it Right}: the low-mass end of the Milky Way globular clusters is flat ($dN/dM =$  constant, $dN/d\log M \propto M$, data from \citealt{harris96}).}
   \label{fig:gcmf}
\end{center}
\end{figure}

With the discovery of young massive clusters (YMCs) in starburst galaxies with masses and radii similar to old GCs  \citep[e.g.][]{holtzman1992, whitmore95} the idea that GCs are still forming today has gained popularity. However,  the mass function of YMCs (power-law with index $-2$, see Fig.~\ref{fig:gcmf}) is strikingly different from the Milky Way GC mass function (GCMF, power-law with index $0$ below a mass of roughly $10^5\msun$).
The GCMF is very similar across many galaxies \citep[e.g.][]{harris14} and this has led to different suggestions for  the origin of the GCMF. On one hand, a near universal `peaked' GCMF could be the result of special conditions in the early Universe causing GCs to form with a preferred mass scale of a few $10^5\,\msun$ \citep[]{fall85, bromm02,kimm15}. However, low-mass clusters dissolve faster in the Galactic tidal field, hence dynamical evolution also plays a role in shaping the GCMF.
Some theoretical studies have claimed that it is possible that GCs formed with the same cluster initial mass function (CIMF) as YMCs and that dynamical evolution can erode the low-mass end of the GCMF by  two-body relaxation  \citep{fall01,kravtsov05, mclaughlin08}, but the required escape rate and its dependence on mass and density are inconsistent with results from $N$-body models of clusters dissolving in a tidal field \citep{vesperini01, gieles08}. There may be additional disruptive mechanisms, such as interactions with left-over natal gas, that could be important in shaping the GCMF soon after formation \citep{kruijssen15}.

We address the relative contribution of nature and nurture to the properties of GCs, and in particular the GCMF, by  using a (inverse) population synthesis approach to include the effect  of a Hubble time of dynamical evolution. Because the dynamical evolution of gravitational systems can only be modelled forward in time, we developed a fast modelling techniques that allows us to efficiently explore the parameter space that describe the initial conditions. 

In  \S\,\ref{sec:simple} we present a theoretical framework for GC evolution, with an application to scaling relations of Milky Way GCs. A fast cluster model and results of a population fitting exercise are presented in \S\,\ref{sec:fit}. Conclusions and a discussion are presented in \S\,\ref{sec:conclusions}.

\section{Model for cluster evolution and a comparison to  Milky Way GCs}
\label{sec:simple}
\subsection{GC evolutionary tracks}

\begin{figure}
\begin{center}
 \includegraphics[width=0.75\columnwidth]{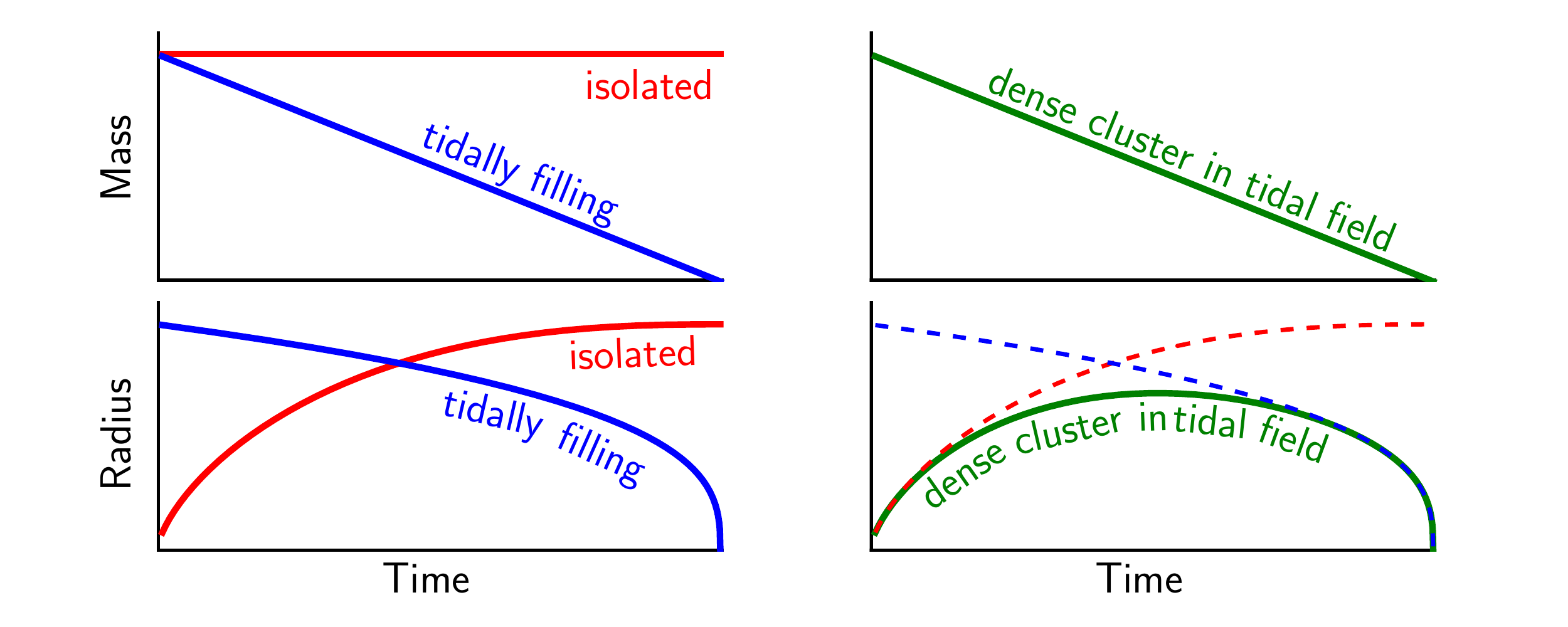} 
 \caption{{\it Left}: Schematic illustration of the evolution of $M$ and $\rh$ in the models for isolated clusters \citep{henon65} and tidally limited clusters \citep{henon61}. {\it Right}: Evolution of the same parameters in a model that `stitches' together the two H\'{e}non solutions \citep*{gieles11}.}
   \label{fig:henon}
\end{center}
\end{figure}

Michel H\'{e}non introduced two models for the evolution of GCs. His models describe the asymptotic behaviour after many relaxation times, and he showed that then  clusters evolve self-similarly.  In both models an  energy source in the core supplies energy at the correct rate to fuel the two-body relaxation process, that is, the energy source is self-regulating and the evolution is `balanced'.
 In the first model \citep{henon61}, the cluster is assumed to be truncated 
 to mimic the effect  of a Galactic tidal field, and stars can escape more easily than from an isolated cluster. The cluster loses stars at a constant rate and evolves at a constant density  that is set by the tidal density, i.e., $\rh\propto (M/\rhotid)^{1/3}$. In the second model \citep{henon65}, the cluster evolves in isolation, and the cluster loses no stars, and $t/\trh$ becomes constant, such that  $\rh\propto t^{2/3}$. Both models are highly idealised, because most GCs are neither in isolated, nor completely filling the tidal volume, but rather somewhere in between these two extremes. It is therefore interested to consider how the two models of H\'{e}non connect and consider the two models  to describe the early and final phases of GC evolution. The fractional energy increase in  H\'{e}non's models is a constant per unit of $\trh$ ($\trh\dot{E}/E =\,$constant) and the constant is approximately the same in the two models \citep{gieles11}. 
This latter property allows us to `stitch'  the two solutions together to acquire a single model that evolves the properties of  clusters in a tidal field. The cluster is initially following the evolution of the isolated model, and while it expands it slowly converges to the constant density solution of the tidally limited model, with the final density set by the strength of the tidal field (see Fig.~\ref{fig:henon}). 

\subsection{GC evolutionary `isochrones'}
\begin{figure}
\begin{center}
 \includegraphics[width=0.9\columnwidth]{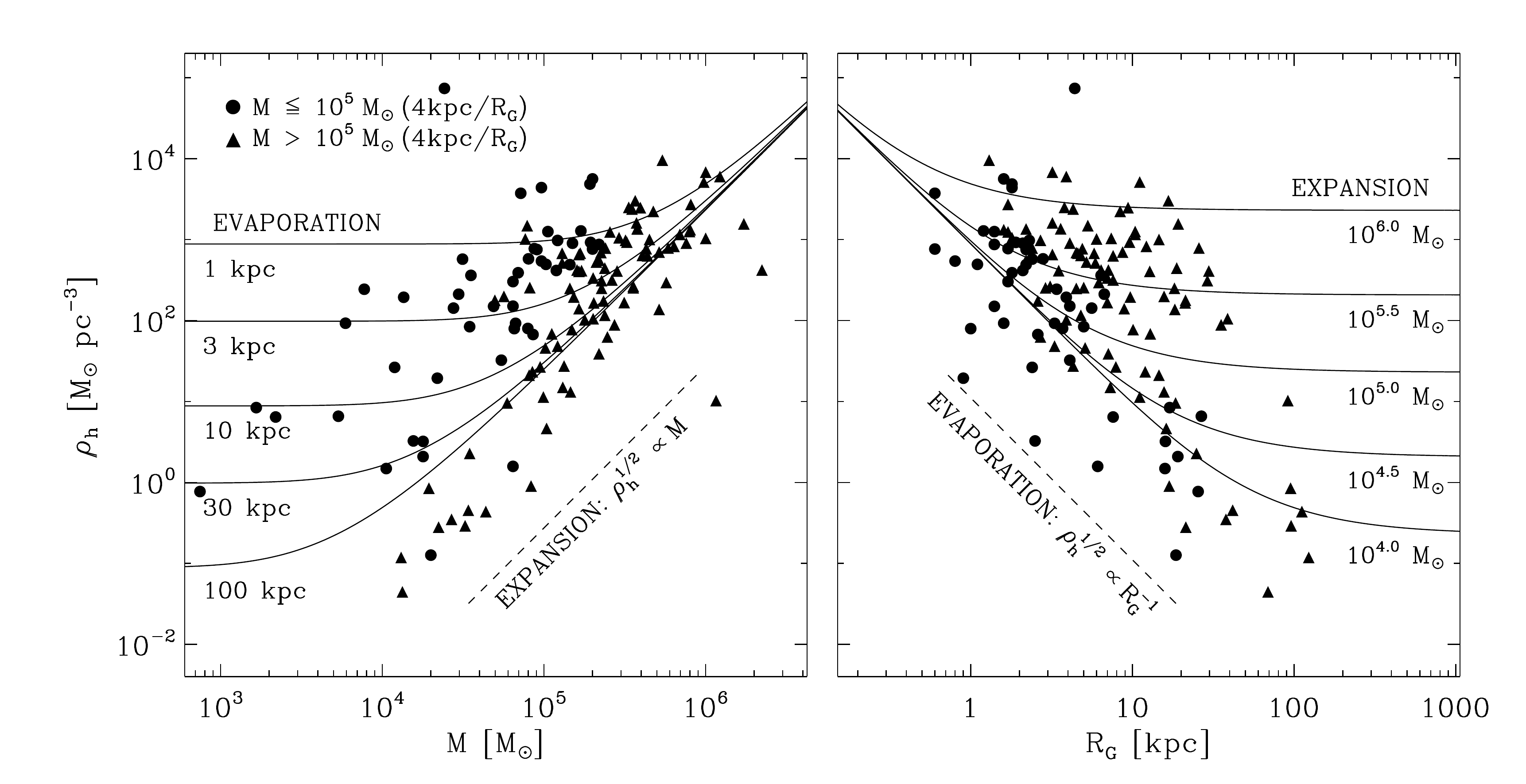} 
 \caption{Evolutionary `isochrones' computed from the model by \citet*{gieles11} overlayed on the properties of Milky Way GCs. {\it Left:} half-mass density $\rhoh = 3M/(8\pi\rh^3)$ vs $M$. The correlation between $\rhoh\propto M^2$ is the result of expansion. {\it Right:} $\rhoh$ vs. $\rg$. The anti-correlation  $\rhoh\propto \rg^{-2}$ is the result of the tidal limit. From this comparison we see that about half of the GCs is still expanding to the tidal boundary and their properties are insensitive of the tidal field they are in. }
   \label{fig:gc_scaling}
\end{center}
\end{figure}

With this simple model for the mass and radius evolution we can also  compute evolutionary `isochrones' for GC properties at an age of 12\,Gyr in a Milky Way-like Galactic potential.
The result is shown in Fig.~\ref{fig:gc_scaling}. It shows the density within $\rh$ ($\rhoh$) as a function of mass $M$ (left) and Galactocentric radius $\rg$ (right), together with the values for the Milky Way GCs. It shows that  about half of the GCs follows the predicted correlation between $\rhoh$ and $M$ of H\'{e}non's isolated cluster (a constant $\trh$, or $\rhoh\propto M^{2}$) and the other half of the GCs follows the prediction of H\'{e}non's  tidally limited cluster (which for the adopted singular isothermal Galactic halo corresponds to a scaling of $\rhoh \propto \rg^{-2}$). 
Because these predicted scaling relations are  asymptotic solutions, they are insensitive to the initial mass-radius relation. All we can conclude from this is thus that GCs must have been denser than they are now. It is possible to gain insight in the initial properties by  considering the {\it distributions} of $M$, $\rhoh$ (or $\rh$) and $\rg$ (i.e. the `Hess diagrams').

\subsection{GC `Hess diagrams'}

\begin{figure}
\begin{center}
 \includegraphics[width=\columnwidth]{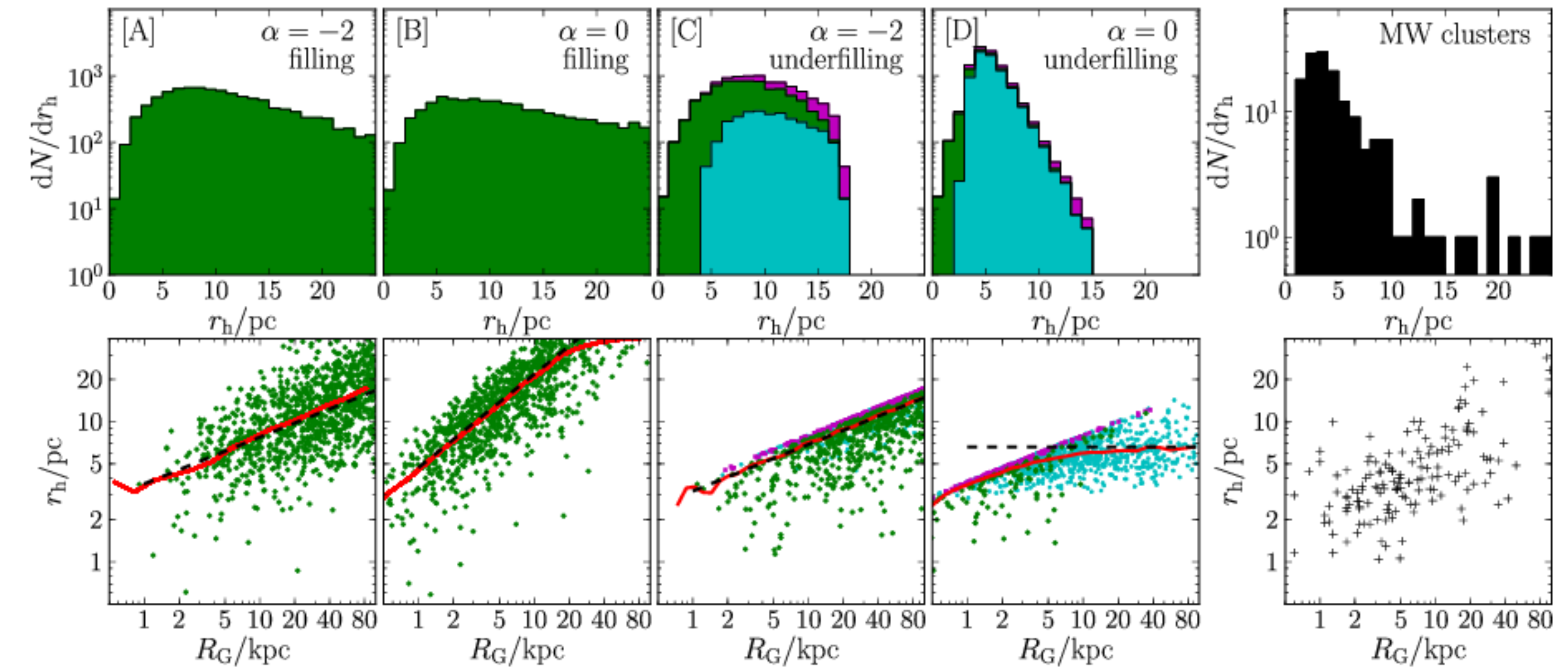} 
 \caption{Radius distribution for different assumptions for the CIMF and the initial $\rh$ distribution (from \citealt{alexander13}) compared to the Milky Way GCs (right panels). Cyan and green marks GCs that are under-filling and filling the Roche radius, respectively, at an age of 12 Gyr and GCs near their maximum $\rh$ are shown in purple. A flat CIMF and dense initial conditions results in a near universal $\rh$ distribution and a weak correlation between $\rh$ and $\rg$.}
   \label{fig:rad}
   \end{center}
\end{figure}

The simple (analytic) prescription for the combined evolution of $\rh$ and $M$ as a function of $\rg$  allows us to explore various scenarios for the birth properties of GCs and compare the evolved  properties to observations.  From data of the ACS Virgo Cluster Survey, it was found that the shape of the $\rh$ distribution across galaxies  is  close to universal \citep{jordan15}.  
Motivated by this finding, we use the model to make predictions for the   $\rh$ distribution of the Milky Way GC population. 
A Monte Carlo approach was used  to simultaneously evolve $M$ and $\rh$ for a large number of clusters for different (extreme)  assumptions for the CIMF and the initial $\rh$ distribution: [A]  tidally filling GCs with a $-2$ power-law CIMF; [B] tidally filling GCs,  with  a flat CIMF ; [C] tidally under-filling GCs, with a $-2$ power-law CIMF; and [D] tidally under-filling GCs, with a flat CIMF. In Fig.~\ref{fig:rad} we show the resulting distributions of $\rh$ for the four scenarios, compared to the $\rh$ distribution of Milky Way GCs (right). Only in the scenario in which the CIMF was flat at birth and all GCs were tidally under-filling,  the narrow shape of the $\rh$ distribution can be reproduced. The peak at $\sim 3-4\,\pc$ is approximately universal, because for about half of the GCs $\rh$ is set by  expansion (i.e. an internal mechanism), and not by the strength of the tidal field. We see that this scenario also results in the correct scaling between $\rh$ and $\rg$.  From a comparison of model [C] and [D], which have the same initial $\rh$ distribution, we see that the resulting $\rh$ distribution is sensitive to the CIMF. The shape of the $\rh$ distribution, therefore, provides additional constraints on the shape of the CIMF.
 
\section{A fast cluster evolution model (\emacss) and an hierarchical Bayesian fit to the Milky Way GCs}
\label{sec:fit}
There are numerous numerical techniques to evolve GC properties, such as solving the Fokker-Planck equations, Monte Carlo modelling and the direct $N$-body integrations. Each of them has its own draw-backs and advantages (see the supplementary material in \citealt{portegies10} for a review), but in general it is the case that the more accurate techniques tend to be slower. 
 \citet{alexander12} developed a method that is comparable to the gas models \citep[e.g.][]{larson70}, in which a coupled set of ordinary differential equations (ODEs) for $\dot{N}$ and $\rhdot$ is solved, using the assumption that $\trh\dot{E}/E$ is constant  during the entire evolution (see \S\,\ref{sec:simple}). In subsequent versions of the model the effect of core evolution \citep{gieles14}, stellar evolution \citep{alexander14} and $\rg$ evolution as the result of dynamical friction were included, resulting in a code that can evolve a limited number of GC properties for the entire lifecycle in few thousand ODE integration steps, and the results show good agreement with results from direct $N$-body simulations\footnote{The code Evolve Me A Cluster of StarS (\emacss) is available from \url{https://github.com/emacss}}. This fast modelling technique allows us to iteratively explore parameter space for the initial conditions of GCs. 
An example of an application of \emacss\ coupled to a Markov Chain Monte Carlo (MCMC) method to find the posterior distribution of the initial parameters of individual GCs can be found  in \citet{pijloo15}.

\begin{figure}
\begin{center}
 \includegraphics[width=0.55\columnwidth]{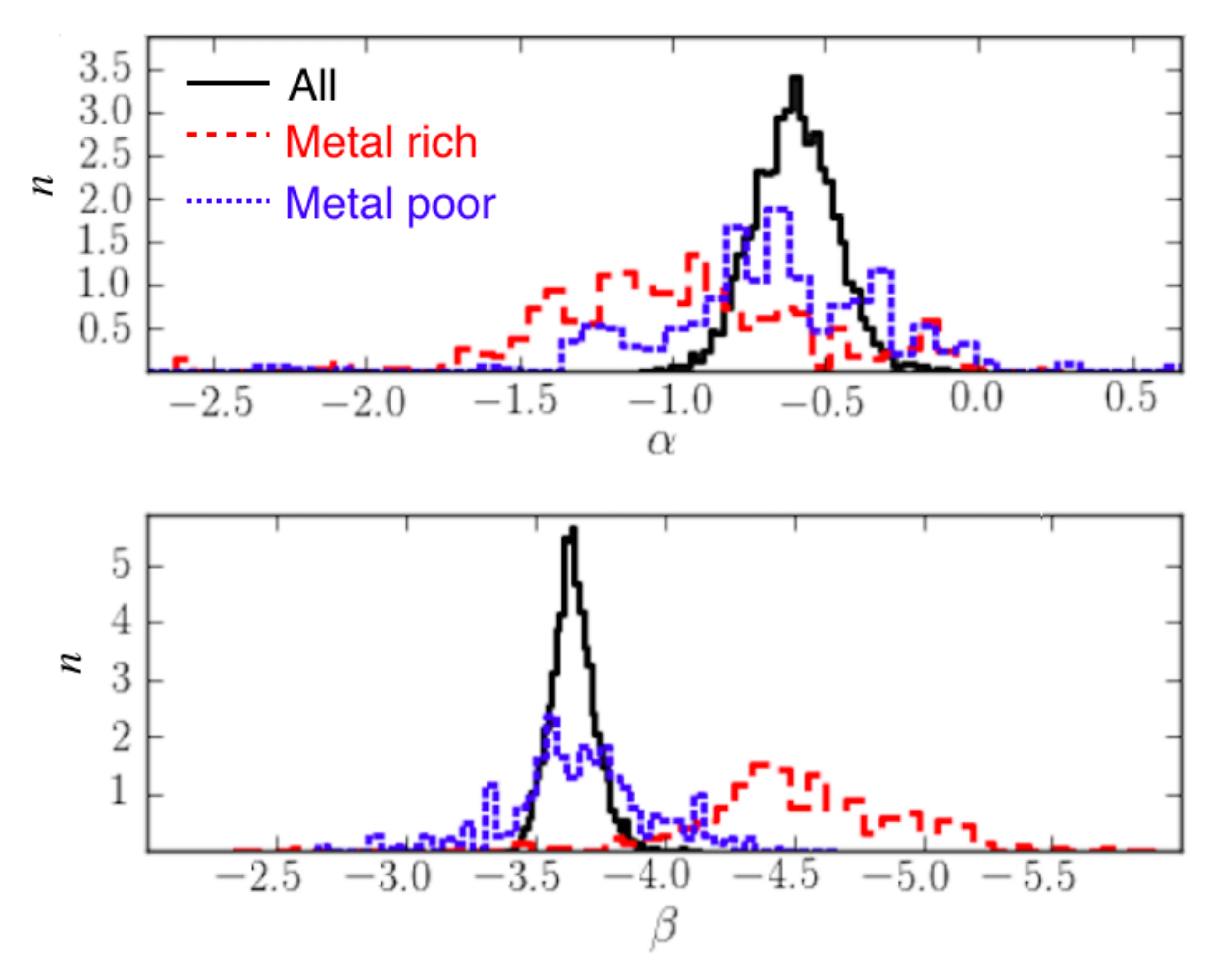} 
 \caption{Posterior distribution for the power-law indices  $\alpha$ of the CIMF  and $\beta$ of the initial number density in the Galaxy for the Milky Way GC population (from Alexander \& Gieles (2106, to be submitted).}
   \label{fig:pop}
\end{center}
\end{figure}

Thanks to the speed of  \emacss, we are able to get the posteriors of every Milky Way GC. We then take this a step further and use this to establish the parameters of the initial {\it distributions} of $M$, $\rh$ and $\rg$ of the Milky Way GCs, with a hierarchical Bayesian approach. The technique is similar to the method presented by \citet{hogg10} to infer the parameters of the eccentricity distribution of planets (or binary stars) from radial velocity measurements. 
We adopt a Schecher function for the CIMF, with parameters $\alpha$ for the index at low masses and $M_*$ for the exponential truncation and a single power-law $\rg^{\beta}$ for the number density distribution of GCs within the Galaxy. Various assumptions for the initial $\rhoh$ distribution have been explored and are discussed in Alexander \& Gieles (2016, to be submitted). 
Our five parameters that describe the initial properties are $\thetapop = (\alpha, M_*, \beta, \murho, \sigmarho)$, where $\murho$ and $\sigmarho$ are the (logarithm of the) mean and the standard deviation of a log-normal for the $\rhoh$ distribution. In the first step we determined the posteriors for the initial parameters of each individual  GC, $P(\thetagc|\gcdat)$, where $\thetagc = (\mi, \rhi, \rgi)_j$ are the parameters of the individual GC and  $\gcdat=(M, \rh, \rg)_j$ are the observed data of  that GC. We use \emacss\ to map the initial properties into present day properties and  in the first iteration we adopt uniform priors for $\mi, \rhi$ and $\rgi$. In the second step we use these posteriors as a set of heteroscedastic observations (i.e. $\popdat = \{P(\thetagc|\gcdat)\}_{j=1}^{\ngc}$) to find the posteriors of the GC population: $P(\thetapop|\popdat)$. We then repeat the first step and use the posteriors $P(\thetapop|\popdat)$  as priors in the fitting of the individual GC parameters $\thetagc$. After three iterations the method converges and we obtain posteriors for $\thetapop$ and $\thetagc$ for each GC. In Fig.~\ref{fig:pop} we show the posterior distributions for $\alpha$ and $\beta$ when considering all GCs, as well as for  the metal-poor and metal-rich GCs separately, where we used $\feh=-1$ as the boundary. The total population is well described by a CIMF with slope of $\alpha \simeq -0.6\pm0.2$, with indications for a slightly steeper distribution for the metal-rich GCs ($\alpha\simeq -1.2\pm 0.4$). The present-day GCMF of metal-rich and metal-poor GCs is very similar, but because of their more  centrally concentrated distribution in the Milky Way (see bottom panel), the metal-rich GCs experience on average a stronger tidal field, such that the present-day GCMF can be reconciled  with a  a slightly steeper CIMF, but within $1-\sigma$ the values of $\alpha$ of the metal-poor and metal-rich GCs are the same.

\section{Conclusions and discussion}
\label{sec:conclusions}
The scaling relations of GCs with masses $\lesssim10^6\msun$ are  the result of  two-body relaxation driven expansion and evaporation in the tidal field. The radii of most GCs are independent of their initial radii, hence we can not rule out that GCs formed with the same $M-\rh$ relation as more massive pressure supported systems, such as UCDs.
About half of the GCs is still expanding to fill their Roche volume and  their $\rh$ are  insensitive to the strength of the tidal field. These expanding clusters have similar $\trh$ (and hence $\rh\propto M^{-1/3}$). The alignment of GCs with lines of constant $\trh$ in a $M-\rh$ diagram has previously  been attributed to the preferential disruption of clusters with short $\trh$ \citep{fall77,gnedin97,mclaughlin08}, but here we show that this scaling is due to expansion, and that GCs that are in the evaporation dominated phase of their evolution (roughly the 2nd half of their life), have adjusted to the tidal density and hence show a relation $\rhoh \propto \rg^{-2}$ (for a host Galaxy with a flat rotation curve), independent of $M$. 
Massive GCs at large $\rg$ are still expanding and because this is an internally driven mechanism (i.e. independent of environment), this provides an explanation for the near universal $\rh$ distribution of extra-galactic GCs.

Previous studies have shown that it is hard to evolve a $-2$ power-law CIMF into a  peaked GCMF with a universal peak mass, when considering stellar evolution and two-body relaxation in a Milky Way-like tidal field \citep[e.g.][]{vesperini01}. We are able to put  constraints on the allowed shapes of the CIMF, given the observations, by using the fast cluster evolution code \emacss\ to evolve parameterised distributions for the initial properties and fit them to the Milky Way GC data. We find that the low-mass end of the CIMF had a slope of $\alpha \simeq -0.6\pm0.2$. A flat CIMF for metal-poor GCs is also inferred from the mass in field stars and GCs in dwarf spheroidals \citep{larsen12, larsen14}. The CIMF of metal-rich GCs in the Milky Way may have been steeper ($\alpha\simeq-1.2\pm0.4$).

The models presented here do not include the disruptive effect of interactions with (molecular) gas clouds, which could be important in the early evolution of GCs \citep{gieles06, kruijssen15}, but several coupled conditions for the initial mass-radius relation and ambient gas density need to be satisfied to establish a near universal GCMF shape. 
 We also only considered a time-independent Milky Way potential. We therefore do not properly account for the effects of the secular growth of the Milky Way and the possibility of GC evolution in a smaller host galaxies that has now been accreted onto the Milky Way, both of which are especially important for the (outer halo) metal-poor GCs. Accounting for the secular growth of the Milky Way results in a higher $\alpha$ (i.e. flatter CIMF), because GCs would have formed further out in a lower-mass Galaxy, and lost less stars than in the models we used to constrain the CIMF \citep{renaud15}. The effect of the evolution in another host galaxy is harder to quantify. \citet{bianchini15} showed that the accretion process itself has little effect on the mass, but whether a period of evolution in a dwarf  galaxy will result in a higher, or larger $\alpha$, depends strongly on the dark matter density profile of the progenitor galaxy. In the $\Lambda$CDM cosmology, smaller galaxies have higher (dark matter) densities, hence we may naively expect the tides to be stronger in dwarf galaxies, compared to the outer halo of the Milky Way. However,  the slope of the density profile is almost as important as the density itself in setting the mass-loss rate.
A fraction of the low-mass satellites appears to/is expected to have dark matter `cores' (i.e. a near constant density) \citep{read15}. Within constant density cores GCs would hardly lose any stars, because of the compressive tidal field. Because these cored satellites are preferentially destroyed when falling into the Milky Way \citep{penarrubia08}, considering the accretion history of GCs within a cosmological context may also result in larger values for $\alpha$ (i.e. flatter CIMF) than reported here. A full consideration of the (tidal) evolution of GCs within the cosmological context will shed light on these issues (Renaud et al., in prep).
\\\\

\end{document}